\documentclass[fleqn,11pt]{wlscirep}
\usepackage[utf8]{inputenc}
\usepackage[T1]{fontenc}
\usepackage{hyperref}
\usepackage{authblk}
\usepackage{xcolor}
\usepackage{ragged2e}

\definecolor{R0}{rgb}{0,0,255}

\title{Probing Non-Equilibrium Grain Boundary Dynamics with XPCS and Domain-Adaptive Machine Learning}

\author[1,2,3,$\dagger$,*]{Mouyang Cheng}
\author[1,4,$\dagger$]{Bowen Yu}
\author[1,5]{Chu-Liang Fu}
\author[6]{Nina Andrejevic}
\author[7]{Matthias T. Agne}
\author[8]{Riley Hanus}
\author[1,5]{Qiwei Wan}
\author[1,9]{Nathan C. Drucker}
\author[1,5]{Thanh Nguyen}
\author[10]{Andrei Fluerasu}
\author[10]{Lutz Wiegart}
\author[10]{Xiaoqian M Chen}
\author[11]{Daniel Pajerowski}
\author[11]{Yongqiang Cheng}
\author[12]{Joshua J Turner}
\author[8]{G. Jeffrey Snyder}
\author[1,2,5,**]{Mingda Li}

\affil[1]{Quantum Measurement Group, MIT, Cambridge, MA 02139, USA}
\affil[2]{Center for Computational Science and Engineering, MIT, Cambridge, MA 02139, USA}
\affil[3]{Department of Materials Science and Engineering, MIT, Cambridge, MA 02139, USA}
\affil[4]{Department of Physics, MIT, Cambridge, MA 02139, USA}
\affil[5]{Department of Nuclear Science and Engineering, MIT, Cambridge, MA 02139, USA}
\affil[6]{Center for Nanoscale Materials, Argonne National Laboratory, Lemont, IL 60439, USA}
\affil[7]{Materials Science Institute, University of Oregon, Eugene, OR 97403, USA}
\affil[8]{Department of Materials Science and Engineering, Northwestern University, Evanston, IL 60208, USA}
\affil[9]{Department of Applied Physics, School of Engineering and Applied Sciences, Harvard University, Cambridge, MA 02138, USA}
\affil[10]{National Synchrotron Light Source II, Brookhaven National Laboratory, Upton, NY 11973, USA}
\affil[11]{Neutron Scattering Division, Oak Ridge National Laboratory, Oak Ridge, TN 37831, USA}
\affil[12]{SLAC National Accelerator Laboratory, Menlo Park, CA 94025, USA}

\affil[$\dagger$]{These authors contributed equally.}
\affil[*]{e-mail: vipandyc@mit.edu}
\affil[**]{e-mail: mingda@mit.edu}

\begin{abstract}
Grain-boundary (GB) dynamics control the stability, mechanical, and functional response of nanocrystalline materials, but direct experimental access to their slow non-equilibrium motion has been limited. Here we establish X-ray photon correlation spectroscopy (XPCS), combined with domain-adaptive machine learning, as a quantitative probe of GB dynamics. Temperature- and grain-size-dependent two-time XPCS measurements in nanocrystalline silicon reveal pronounced departures from time-translation invariance, showing that GB relaxation can remain far from equilibrium over experimental timescales. However, direct extraction of quantitative physical information from these high-dimensional, noisy fluctuation maps faces a significant challenge. To overcome this barrier, we develop a semi-supervised learning framework that transfers physical parameter labels from continuum simulations to unlabeled experimental XPCS maps through domain-adaptive representation alignment. This AI-augmented approach enables the extraction of key kinetic parameters, including bulk diffusivity, GB stiffness, and effective GB concentration, directly from experimental XPCS measurements. Our results show how machine learning can transform indirect fluctuation signals into quantitative materials dynamics, providing a general route to study non-equilibrium defect motion in solids.
\end{abstract}

\begin{document}

\flushbottom
\maketitle
\thispagestyle{empty}

\section*{Introduction}
Nanograined materials derive many of their remarkable properties from the dense grain-boundary (GB) networks that thread through their microstructure. In these materials, GBs are not static interfaces, but dynamic mesoscale objects whose migration, sliding, relaxation, and structural transformation can reshape materials over time. Stress-driven GB migration shows that GBs can move under shear rather than simply obstructing dislocations \cite{rupert2009experimental}, while atomic-resolution studies reveal coupled sliding, atomic transfer, disconnection motion, and structural transformation during GB deformation \cite{wang2022tracking,fang2022atomic}. In nanograined alloys, relaxation and stabilization of high-density GB networks can suppress diffusional creep and sustain high-temperature strength \cite{zhang2022inhibiting}; under irradiation, GB facet motion and interfacial defect dynamics govern microstructural evolution \cite{barr2022irradiation}. Thus, nanograined solids are governed not only by GB structure, but by GB dynamics: how boundaries move, fluctuate, relax, and transform under external perturbations.

Despite this importance, experimental access to slow GB dynamics remains limited. Many relevant relaxation modes occur over minutes to hours and manifest as weak, spatially correlated fluctuations rather than abrupt morphological changes. High-energy diffraction microscopy can track grain-scale evolution \cite{bernier2020high}, while four-dimensional electron microscopy \cite{zewail2010four,tian2024grain}, Bragg coherent diffraction imaging \cite{robinson2016materials,yau2017bragg}, and time-resolved X-ray tomography \cite{garcia2021tomoscopy,mckenna2014grain,zhang2020grain} provide complementary views of structure, strain, and mesoscale morphology. Yet none readily combines the temporal window, spatial sensitivity, and statistical averaging needed to quantify slow, non-equilibrium GB relaxation. Consequently, key dynamical laws, including curvature-driven migration, stress-assisted motion, thermal activation, impurity pinning, and noise-activated hopping, remain difficult to validate experimentally despite extensive theory \cite{gottstein2005triple,upmanyu1999misorientation,taylor1992overview,mishin1997atomistic,kobayashi2000continuum,trautt2012grain}.

In this work, we show that X-ray photon correlation spectroscopy (XPCS), when augmented by specialized machine learning, can directly reveal GB dynamics in nanocrystalline materials. XPCS has long been used to probe time-resolved nanoscale dynamics in soft matter, glasses, colloids, and other complex systems \cite{leheny2012xpcs,bikondoa2016x,lu2010temperature,zinn2020phoretic}. 
More recently, it has also been extended to probe slow fluctuations of ordered states in hard condensed matter systems \cite{porter2024understanding,shen2023interplay,tumbleson2025thermodynamic}.
However, extracting quantitative physical information from XPCS remains highly challenging because the measured two-time correlation maps are high-dimensional, noisy, and often far from equilibrium. Recent machine-learning approaches have enabled automated classification of relaxation dynamics from experimental XPCS data \cite{horwath2024ai}, but converting such measurements into material-specific kinetic parameters remains largely inaccessible.

Here, we overcome this challenge by designing a semi-supervised domain-adaptive learning framework that bridges continuum simulations and experimental XPCS measurements. Our model uses simulated GB dynamics to learn the relationship between two-time correlations and physical parameters, while experimental data constrain the learned representation through domain alignment and non-equilibrium consistency. This enables, for the first time, direct extraction of key GB-motion parameters, including bulk diffusivity $D$, GB stiffness $\Gamma$, and effective GB concentration $\lambda_\mathrm{GB}$, from experimental XPCS maps. It also allows us to directly quantify how far nanograined silicon departs from equilibrium through the breakdown of time-translation invariance. By turning XPCS into a quantitative measurement of GB dynamics, our work establishes a new route for studying nanograined materials across long timescales and weak fluctuation regimes inaccessible to real-space imaging. More broadly, the framework provides a generally applicable strategy for synergizing experimental and computational data, marking a step toward applying AI directly to noisy, high-dimensional experimental observables in complex materials systems.

\section*{Results}
\subsection*{Workflow}
\begin{figure}[!htbp]
  \centering
  \includegraphics[width=\textwidth]{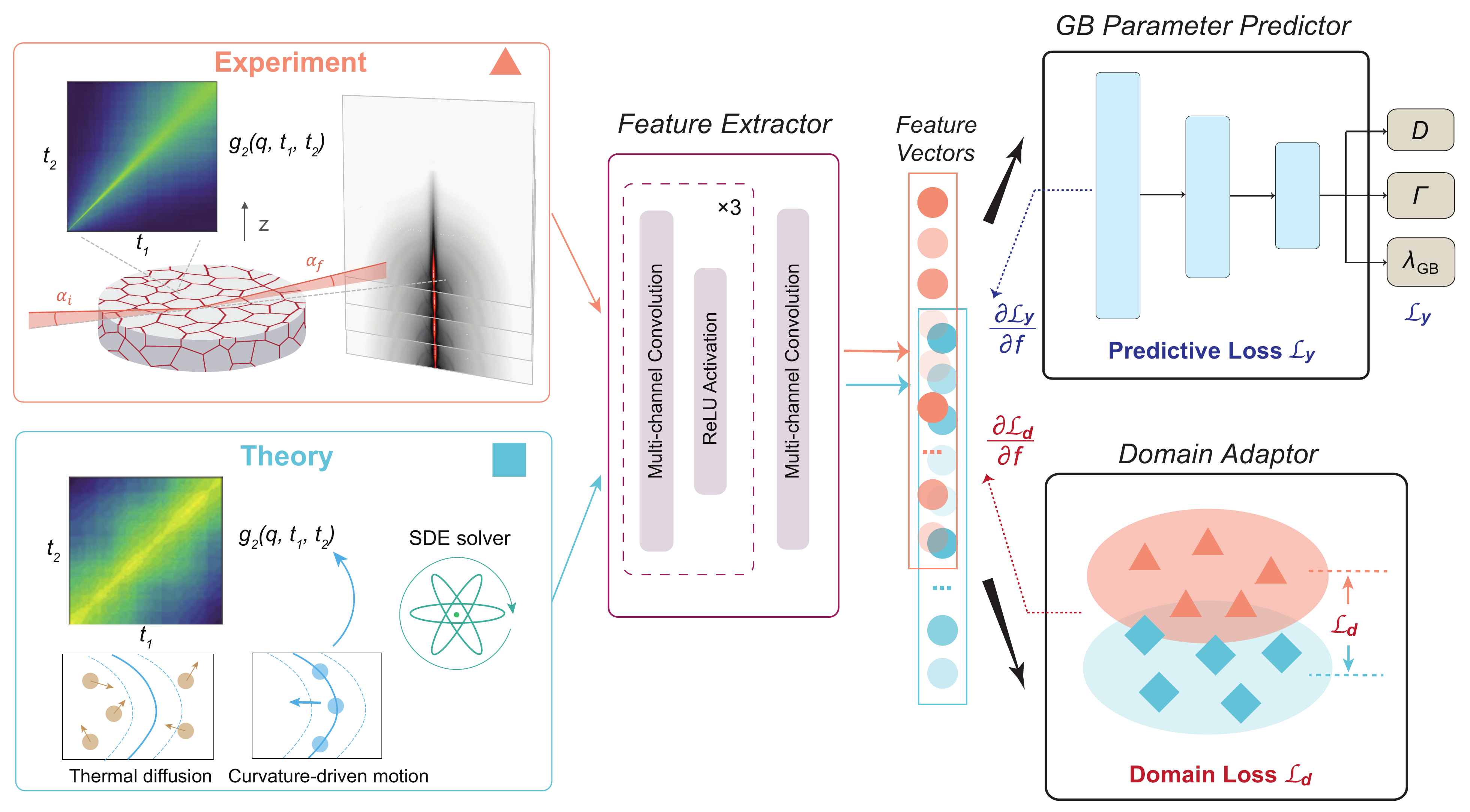}
  \caption{\textbf{Integrated workflow with experiment, theory, and AI for grain-boundary dynamics.} Experimentally, X-ray photon correlation spectroscopy (XPCS) measures the two-time correlation function $g_2(\mathbf{q},t_1,t_2)$, capturing non-equilibrium dynamics of the grain boundary.
  Theoretically, stochastic differential equation (SDE)-based simulations incorporate thermal diffusion and curvature-driven grain-boundary migration to generate theoretical $g_2(\mathbf{q},t_1,t_2)$ maps. To bridge the domain gap between theory and experiment, we employ a semi-supervised domain adaptation framework with two coupled losses: a regression loss $\mathcal{L}_y$ that trains theoretical $g_2$ maps to predict $(D,\Gamma,\lambda_{\mathrm{GB}})$, and a domain alignment loss $\mathcal{L}_d$ that constrains experimental (orange triangles) and theoretical (blue diamonds) embeddings with matched non-equilibrium measures into a shared feature space. This alignment enables direct and robust inference of GB kinetic parameters from experimental XPCS measurements.}
  \label{fig1}
\end{figure}

We integrate experiment, theory, and AI to extract physically meaningful parameters from non-equilibrium XPCS measurements of GB motion, as is illustrated in Fig. \ref{fig1}. 
\textit{Experimentally}, XPCS in a grazing-incidence small angle X-ray scattering  (GISAXS) geometry probes near-surface mesoscopic structures by measuring coherent X-ray speckle patterns from the disordered GB network. 
Correlating these speckle patterns over time reveals the dynamics of the scatterers, yielding the two-time intensity-intensity correlation maps $g_2(\mathbf{q},t_1,t_2)$ across temperatures.
\textit{Theoretically}, the two-time correlation $g_2$ can be computed from atomic-scale dynamics, although directly resolving GB dynamics at that scale could be prohibitively expensive. Therefore, instead of tracking the dynamical trajectory of each silicon atom, we adopt a coarse-grained model, in which GB motion is described by a stochastic differential equation (SDE) for the interfacial height $z(x,t)$, with curvature-driven migration and thermal fluctuations governing its time evolution.
Through this coarse-grained model, one can characterize the GB motion by a few controllable parameters, including the bulk diffusivity $D$, GB stiffness $\Gamma$ and the effective GB concentration $\lambda_\text{GB}$ that contributes to the non-equilibrium motion.
From these trajectories we compute synthetic $g_2(\mathbf{q}, t_1,t_2)$ maps and their non-equilibrium measure, $\mathcal{S}_{\mathrm{noneq}}$ (see definition in Eq. \ref{eq:Non-eq}) over a grid of parameters ($D$, $\Gamma$, $\lambda_{\mathrm{GB}}$). 
Results show that, across a reasonable range of model parameters, the theory reproduces key experimental trends, including the $g_2$ patterns and the crossover from equilibrium to non-equilibrium behavior.

With experimental data and the theoretical model at hand, we aim to assign the theoretical parameters to experimental XPCS data. A natural thought is to train a neural network on simulated $g_2$ maps, use them as image-like input and optimize a regression loss $\mathcal{L}_y$ to predict ($D$, $\Gamma$, $\lambda_{\mathrm{GB}}$). 
However, the domain gap between idealized simulations and experimental measurements, which could arise from background noise, instrumental response, and unmodeled microstructure, prevents direct application of such a model. This is further complicated by the fact that experimental data are scarce and unlabeled, making purely supervised regression infeasible.
To address this, we adopt a semi-supervised learning framework, where both labelled simulated data and unlabelled experimental data are fully utilized. The net training objective is to minimize both the predictive loss $\mathcal{L}_y$ from the GB parameter predictor and the domain alignment loss $\mathcal{L}_d$ from the domain adaptor. 
Both experimental (orange) and simulated (blue) inputs are passed through the same \textit{feature extractor} (with shared weights) to obtain their \textit{feature vectors}. The regression loss $\mathcal{L}_y$ is applied only to simulated data, where ground-truth parameters are available. 
In parallel, a domain alignment loss $\mathcal{L}_d$ encourages the embeddings of simulated and experimental data to overlap, effectively bridging the domain gap.  
As a result, the model learns a shared representation that aligns experimental and theoretical data while preserving accurate parameter estimation from simulation. This enables direct and reliable inference of GB kinetic parameters from experimental XPCS measurements, effectively using AI to bridge theory and experiment within a unified framework.
More details of the training setup are shown in the Methods section.

\subsection*{XPCS measurement}
\begin{figure}[t]
  \centering
  \includegraphics[width=1\textwidth]{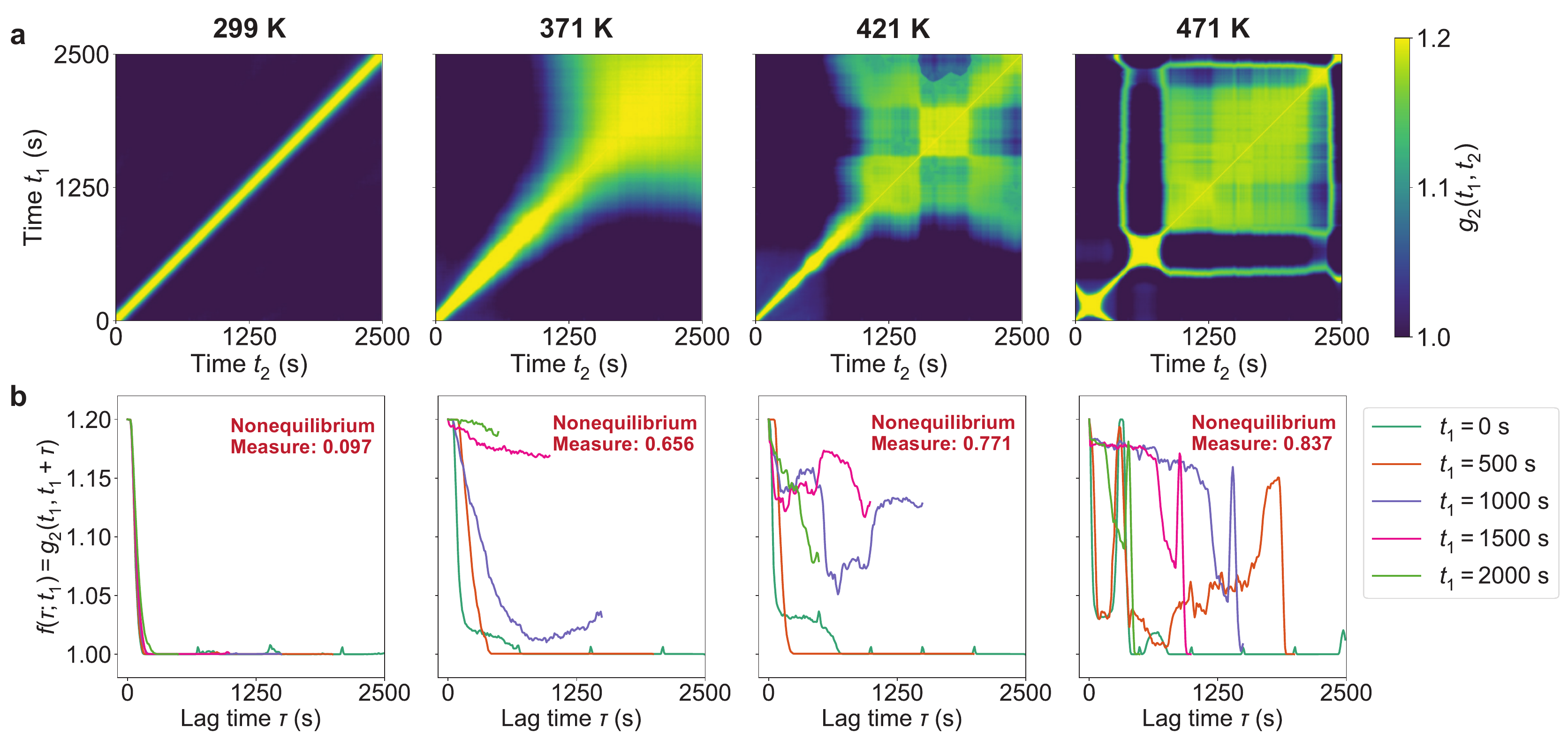}
  \caption{\textbf{Experimental temperature-dependent XPCS dynamics and non-equilibrium behavior.} \textbf{a.} Two-time correlation functions $g_2(t_1, t_2)$ measured at 299 K, 371 K, 421 K, and 471 K for the same nano-crystalline silicon sample. 
  The results reveal increasingly non-equilibrium features in the dynamics as temperature rises. Here, $g_2(t_1,t_2)$ denotes $g_2(\mathbf{q},t_1,t_2)$ evaluated with in-plane momentum transfer $q_r=0.045 \rm{\mathring{A}}^{-1}$ and out-of-plane $q_z=0.275 \rm{\mathring{A}}^{-1}$. \textbf{b.} Corresponding intensity-intensity correlation slices $f(\tau; t_1) = g_2(t_1, t_1+\tau)$ at different times $t_1$. As temperature increases, $f(\tau; t_1)$ demonstrates systematic deviations between different $t_1$, indicating a transition from near-equilibrium to non-equilibrium GB dynamics. A more quantitative non-equilibrium measure following the definition in Eq.\,\ref{eq:Non-eq} is also shown for each temperature.}
  \label{fig2}
\end{figure}

The XPCS is measured for nanostructured bulk silicon, which is synthesized according to the methods described by Ref.\,\cite{bux2009nanostructured}. More details about the XPCS measurements are shown in Methods.
Several representative XPCS measurements of the two-time correlation function $g_2(t_1, t_2)$, acquired over an observation window $0 \le t_1, t_2 \le \mathcal{T} = 2,500~\mathrm{s}$ at selected temperatures $T$ on the same sample, are shown in Fig.\,\ref{fig2}a. Here, $g_2(t_1, t_2)$ denotes $g_2(\mathbf{q}, t_1, t_2)$ evaluated at an in-plane momentum transfer $q_r = 0.045~\mathrm{\mathring{A}}^{-1}$ and an out-of-plane component $q_z = 0.275~\mathrm{\mathring{A}}^{-1}$, with the full scattering vector $\mathbf{q}$ omitted for brevity.
At 299 K, the $g_2$ map is close to time-translation invariant (TTI): intensity is organized mainly along the $\tau=t_2-t_1$ diagonals and is consistent with near-equilibrium dynamics, where $g_2(t_1,t_2)$ may be approximated as $g_2(\tau)$.
Upon heating, this diagonal organization is progressively lost. At 371 K, the ridge broadens and bends into extended off-diagonal regions for $t_1, t_2 \ge 1,250$ s, demonstrating a shift from near-equilibrium dynamics towards non-equilibrium. At 421 K and 471 K, the maps develop complex, highly non-equilibrium structures, including pronounced block-like and banded wedges, which reveals strongly history-dependent, non-equilibrium structural dynamics.
In particular, the box-like structures at 471 K suggest intermittent non-stationary dynamics, in which the system remains correlated within a finite time window and then decorrelates sharply across rearrangement events.

To quantify the non-equilibrium measure for each sample, we utilize the TTI property for thermal-equilibrated dynamics, where $g_2$ depends only on the lag $\tau=t_2-t_1$. Deviations from TTI therefore provide a direct, model-agnostic measure of non-equilibrium. The best TTI ``baseline'' is first obtained by averaging $g_2$ along lines of constant $\tau$ (the diagonals of the two-time plane as shown in Fig.\,\ref{fig2}a). Denoting $g_2(t_1,t_2)=g_2(t_1,t_1+\tau)$, we define
\begin{equation}
    g_2^{\mathrm{TTI}}(\tau)\equiv
\frac{1}{N(\tau)}\!
\int_{\max(0,-\tau)}^{\min(\mathcal{T},\mathcal{T}-\tau)}
g_2(t,\,t+\tau)\,dt,
\quad
N(\tau)=\mathcal{T}-|\tau|
\end{equation}
which performs the diagonal average along the $g_2$ map within the appropriate finite-window limits. The non-equilibrium drift $S_{\mathrm{noneq}}(t_1,t_2)$ for each time pair $(t_1, t_2)$ is then defined as
\begin{equation}
    S_{\mathrm{noneq}}(t_1,t_2)= g_2(t_1,t_2)-g_2^{\mathrm{TTI}}(t_2-t_1)
\end{equation}
By construction, $S_{\mathrm{noneq}}(t_1,t_2) \equiv 0$ if the dynamics is fully equilibrated. Building upon this, we can further define the global non-equilibrium measure $\mathcal{S}_{\mathrm{noneq}}$ as
\begin{equation}
    \label{eq:Non-eq}
    \mathcal{S}_{\mathrm{noneq}}= \frac{\big\langle S_{\mathrm{noneq}}(t_1,t_2)^2\big\rangle_{t_1,t_2}}{\big\langle \big(g_2(t_1,t_2) -\big\langle g_2\big\rangle_{t_1,t_2}\big)^2\big\rangle_{t_1,t_2}} \in [0,1]
\end{equation}
i.e., the fraction of the variance of $g_2$ that arises from non-equilibrium effects. The brackets $\langle\cdot\rangle_{t_1,t_2}$ denote time averaging over all $(t_1,t_2)$ pairs within the observation window. 
$\mathcal{S}_{\mathrm{noneq}}$ is naturally normalized between 0 and 1: it is 0 for fully equilibrated dynamics, grows with the strength of non-equilibrium, and is capped at 1 where $g_2$ contains no time-translation-invariant component $\tau$ and all variance arises from explicit $(t_1,t_2)$ dependence.
For better visualization of non-equilibrium effects, we plot $f(\tau; t_1)=g_2(t_1, t_1+\tau)$ as functions of $\tau$ for different $t_1$ in Fig.\,\ref{fig2}b. For the spectrum close to TTI at 299 K, the correlation slices $f(\tau;t_1)$ nearly coincides as a single decay, consistent with a low non-equilibrium measure $\mathcal{S}_{\mathrm{noneq}}=0.097$. For the highly non-equilibrated spectra at 421 K and 471 K, correlation slices $f(\tau; t_1)$ separate significantly according to $t_1$, which correspond to their large non-equilibrium measures $\mathcal{S}_{\mathrm{noneq}} = 0.771$ and $\mathcal{S}_{\mathrm{noneq}} = 0.837$.

\subsection*{Simulation of XPCS signal from GB motion}
To gain physical insight into the non-equilibrium dynamics observed experimentally, we develop a theoretical model where we attribute the leading contribution to GB dynamics. This assumption is justified by the long timescales involved (> 1,000 s, see Fig.\,\ref{fig2}a), which are naturally associated with processes occurring over large spatial scales, such as GB motion.
We start from the Siegert relation \cite{siegert1943fluctuations,berne2000dynamic,bikondoa2016x}, approximating $g_2$ as
\begin{equation}
    g_2(\mathbf{q},t,t') = 1 + \beta |g_1(\mathbf{q},t,t')|^2
\label{g2}
\end{equation}
where $\beta$ is an experiment-dependent constant and  $g_1(\mathbf{q},t,t') = \frac{1}{N} \sum_{i,j} \langle e^{i\mathbf{q}\cdot[\mathbf{r}_i(t)-\mathbf{r}_j(t')]} \rangle$ is the coherent intermediate scattering function, with $N$ the total number of atoms, $\mathbf{r}_i(t)$ the position of atom $i$ at time $t$, and $\langle \cdot \rangle$ denoting an ensemble average.
The intermediate scattering function $g_1(\mathbf{q},t,t')$ can, in principle, be computed directly from atomistic simulations such as molecular dynamics (MD) \cite{allen2017computer}. However, MD is limited to timescales on the order of nanoseconds, far shorter than the microsecond-to-second dynamics probed by XPCS, making a direct comparison infeasible. To facilitate direct comparison with experimental data, we find it sufficient to adopt a continuum model with a mixture of non-equilibrium GB motion and equilibrium diffusion.
Specifically, we split the atoms in the system into two parts: I. the atoms proximal to the grain boundary (GB); II. the atoms away from the GB. We assume that all atoms not on the GB are at equilibrium and the non-equilibrium XPCS signals are mainly attributed to the motion of GB, where the GB configuration is not fully relaxed, possible due to low annealing temperature. Neglecting the cross terms, we approximate $g_1(\mathbf{q},t,t')$ as \begin{equation}
    g_1(\mathbf{q},t,t')/g_0 = \lambda_\mathrm{GB}\left<e^{i\mathbf{q}\cdot (\mathbf{r}_i(t)-\mathbf{r}_j(t'))}\right>_{\mathrm{GB}} + (1-\lambda_\mathrm{GB})\left<e^{i\mathbf{q}\cdot (\mathbf{r}_i(t)-\mathbf{r}_j(t'))}\right>_{\mathrm{eq}}
\end{equation}
Here, $g_0$ is a normalization constant and $\lambda_\mathrm{GB} \in[0,1]$ is the effective ratio of atoms proximal to the GB to the total atoms. Larger grain size means lower concentration of GB, thereby smaller $\lambda_\mathrm{GB}$. 
The equilibrium dynamics are assumed to be governed by fast diffusive motion, and we approximate the corresponding decorrelation by a Brownian form,
\begin{equation}
\left< e^{i\mathbf{q}\cdot \left(\mathbf{r}_i(t)-\mathbf{r}_j(t')\right)}\right>_{\mathrm{eq}}= \left\langle e^{i \mathbf{q} \cdot\left(\mathbf{r}_i(t)-\mathbf{r}_j(t)\right)}\right\rangle e^{-\mathbf{q}^2 D(t'-t)} = S(\mathbf{q})e^{-\mathbf{q}^2 D(t'-t)}, \quad t'>t.
\label{equilibrium}
\end{equation}
where $S(\mathbf{q})$ is the structure factor of the sample, and $D$ is the bulk diffusivity of the atoms in the sample. For a given momentum transfer $\mathbf{q}$, $S(\mathbf{q})$ is time-independent over the XPCS observation window, and it can be absorbed as a prefactor into $\lambda_\text{GB}$ as well.

\begin{figure}[!htbp]
  \centering
  \includegraphics[width=0.9\textwidth]{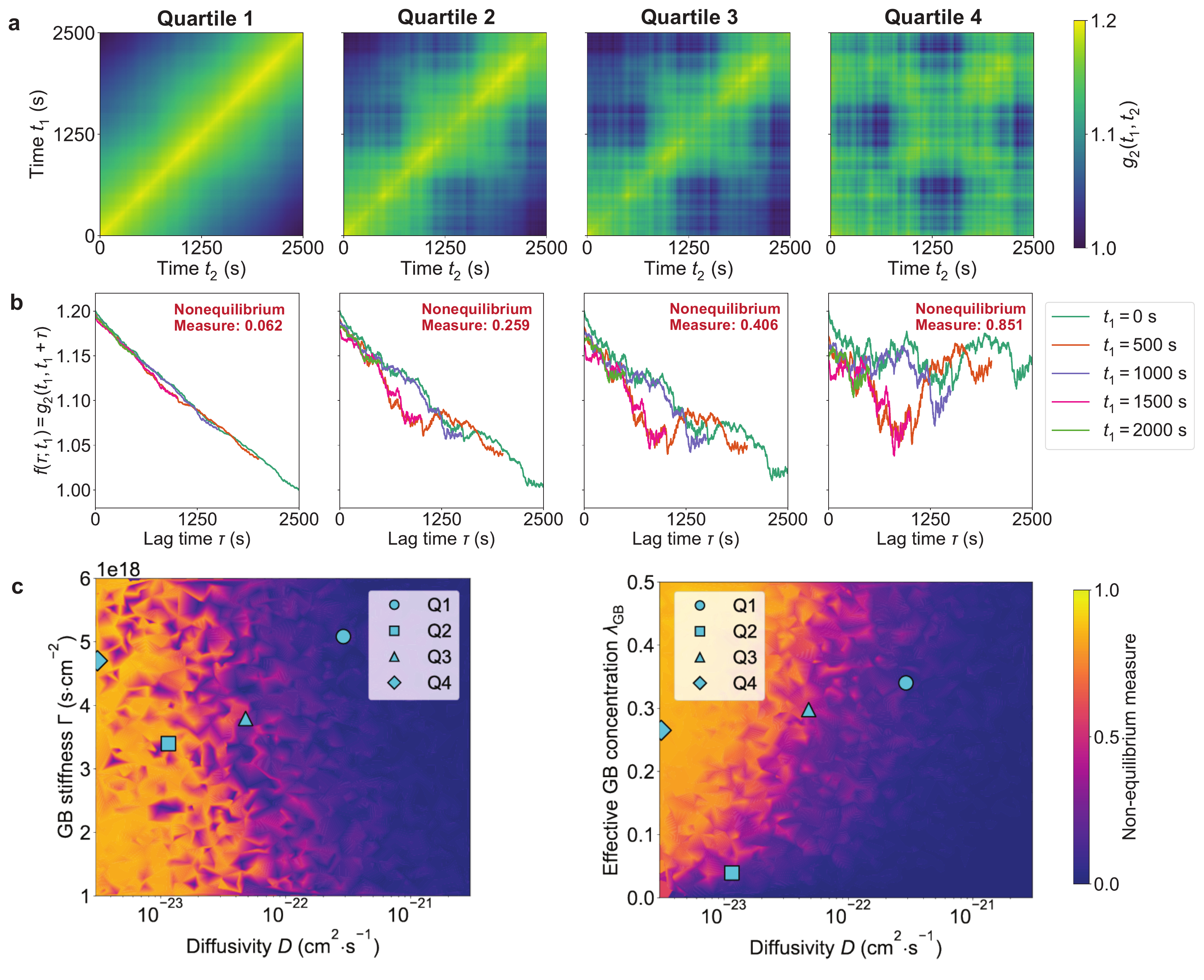}
  \caption{\textbf{Theoretical SDE predictions and mapping of non-equilibrium dynamics across parameter space.}
    \textbf{a.} Simulated two-time correlation functions $g_2(t_1, t_2)$ for four representative parameter sets, showing the progression from nearly time-translation-invariant dynamics to strongly non-equilibrium behavior.
    Here $g_2(t_1,t_2)$ denotes $g_2(\mathbf{q},t_1,t_2)$ with in-plane $q_r=0.045 \rm{\mathring{A}}^{-1}$.
    \textbf{b.} Corresponding intensity-intensity correlation slices $f(\tau; t_1) = g_2(t_1, t_1+\tau)$ at different starting times $t_1$, with non-equilibrium measures $\mathcal{S}_{\mathrm{noneq}}=0.062$, 0.259, 0.406, and 0.851 for the four cases in \textbf{a}. 
    \textbf{c.} Heatmap visualization of the predicted non-equilibrium measure across pairwise parameter spaces involving bulk diffusivity $D$, GB stiffness $\Gamma$, and effective GB concentration $\lambda_\mathrm{GB}$; markers indicate the four representative cases shown in \textbf{a} and \textbf{b}.}
    \label{fig3}
\end{figure}

To compute the contribution from GB motion, we assume that the grain boundary (GB) lies in the $x$-$y$ plane, with out-of-plane deviations described by a displacement field $z=z(x)$, i.e. the morphology varies only along the x-direction and is uniform along y. Thus, the classical equation of motion (EOM) for the GB in the continuum limit\cite{chen2018atomistic} is reduced to a stochastic differential equation (SDE):
\begin{equation}
v(x, t)=\frac{\partial z(x, t)}{\partial t}=M\left[\Gamma \frac{\partial^2 z(x,t)}{\partial x^2}+\xi(x, t)\right]
\label{EOM}
\end{equation}
where $v(x,t)$ is the $z$-directional velocity, $M$ is the mobility constant of the GB, $\Gamma$ is the GB stiffness, and $\xi(x,t)$ denotes the thermal noise. With the EOM and proper setup of initial and boundary conditions, one can numerically simulate the time-evolution of GB from the SDE and finally obtain the XPCS two-time correlation $g_2(t_1, t_2)$, allowing us to study its dependence on physical parameters such as $D$, $\Gamma$ and $\lambda_{GB}$. More details regarding the model and SDE setup for GB dynamics are shown in Methods.

Results for the continuum model simulation are shown in Fig.\,\ref{fig3}, where $g_2(t_1,t_2)$ denotes $g_2(\mathbf{q},t_1,t_2)$ evaluated at an in-plane momentum transfer $q_r = 0.045~\mathrm{\mathring{A}}^{-1}$, similar to Fig.\,\ref{fig2}. To test whether this minimal model can generate the hierarchy of two-time structures seen experimentally, we simulated 2,000 synthetic XPCS maps over a broad kinetic parameter space (Methods). The resulting ensemble spans the entire non-equilibrium measure range from $\mathcal{S}_\text{noneq}=1.8\times10^{-4}$ to $\mathcal{S}_\text{noneq}=0.889$, which also fully covers the experimental range in Fig.\,\ref{fig2}. The four maps in Fig.\,\ref{fig3}a,b were selected from successive quartiles of this distribution and show a continuous evolution from nearly stationary dynamics to the block-like and wedge-shaped correlations characteristic of strongly non-equilibrium relaxation, highly similar to the experimental observation in Fig.\,\ref{fig2}a. 

Furthermore, Fig.\,\ref{fig3}c presents a systematic mapping of the non-equilibrium measure across the parameter space $(D,\Gamma,\lambda_{\mathrm{GB}})$, revealing the interplay between bulk diffusive relaxation, GB curvature-driven dynamics, and effective GB concentration. 
Physically, the bulk diffusivity $D$ governs the rate of equilibrium decorrelation, effectively driving the system towards time-translation invariance. In contrast, the grain-boundary stiffness $\Gamma$ controls the strength of curvature-driven restoring forces, thereby controlling the intrinsic relaxation timescale of GB fluctuations. The effective GB concentration $\lambda_{\mathrm{GB}}$ determines the relative weight of non-equilibrium GB motion in the total scattering signal. As a result, non-equilibrium behavior emerges from a competition between fast bulk diffusion and slow, curvature-driven GB dynamics, with larger $\lambda_{\mathrm{GB}}$ amplifying the contribution of the latter. 
From Fig.\,\ref{fig3}c, we find that decreasing $D$ or increasing $\lambda_{\mathrm{GB}}$ enhances the persistence of non-stationary features in $g_2(t_1,t_2)$ (right panel), while smaller $\Gamma$ further promotes non-equilibrium behavior by weakening curvature-driven restoring forces and slowing GB relaxation (left panel). The parameter map also reveals crossover regimes in both $(\Gamma, D)$ and $(\lambda_{\mathrm{GB}}, D)$ spaces, separating regimes dominated by either diffusive equilibration or GB-driven dynamics. 
Additionally, we highlight in Fig.\,\ref{fig3}c the region of parameter space (Q1-Q4) used in Fig.\,\ref{fig3}a, illustrating how the model can be tuned from equilibrium to non-equilibrium regimes and how parameters can be sampled accordingly for batch simulations.

\subsection*{Bridging theoretical and experimental XPCS with semi-supervised learning}
\begin{figure}[!htbp]
  \centering
  \includegraphics[width=0.9\textwidth]{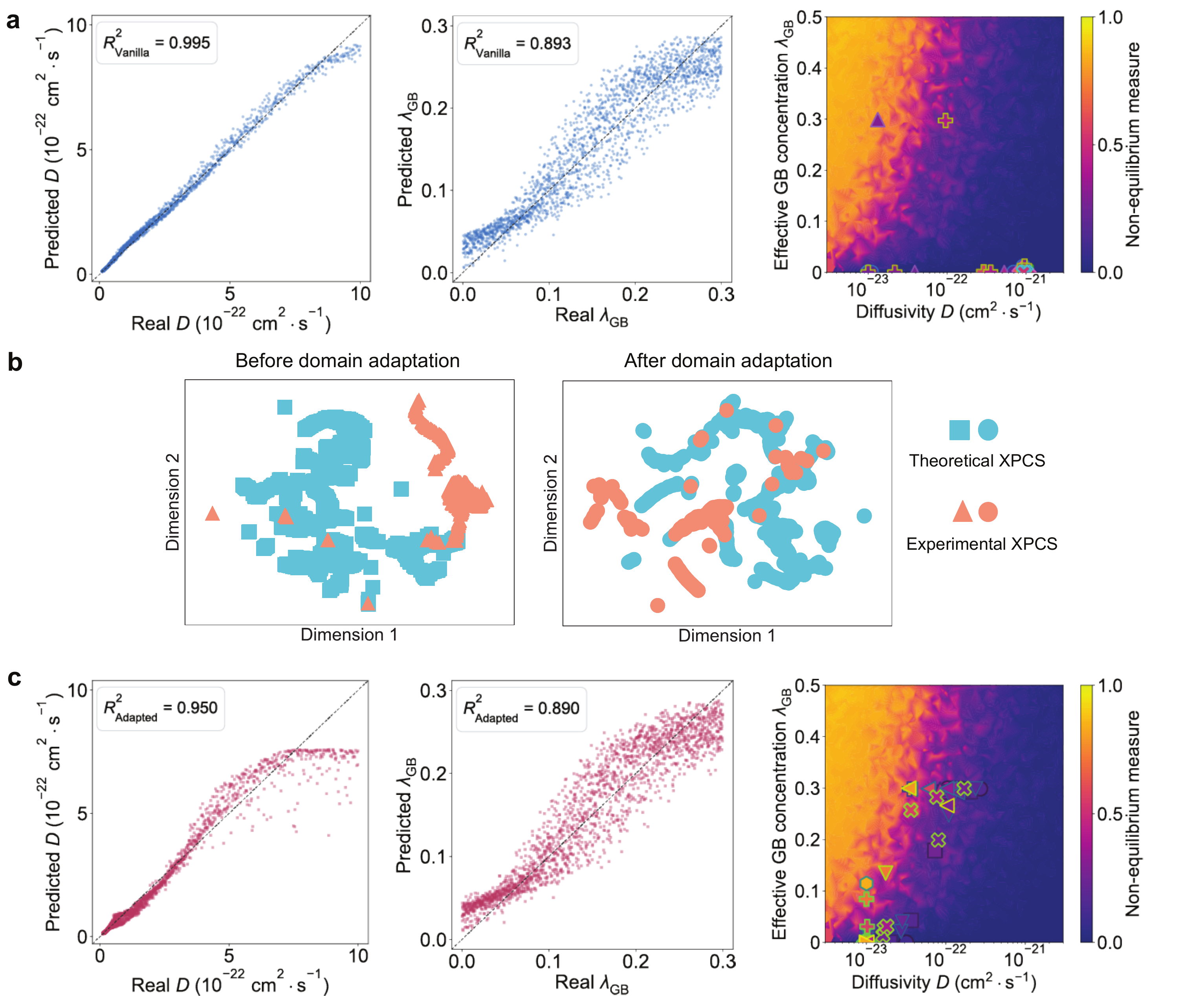}
  \caption{\textbf{Semi-supervised domain adaptation for mapping experimental XPCS onto the theoretical parameter manifold.}
  \textbf{a.} Prediction results before domain adaptation using the vanilla model trained only on simulated XPCS. Left and middle panel: predicted versus real $D$ and $\lambda_{\mathrm{GB}}$ for simulated test data, with a test-set $R^2=0.995$ and $0.893$, respectively. Right panel: inferred experimental XPCS locations overlaid on the theoretical $(D,\lambda_{\mathrm{GB}})$ parameter map, where the background colormap shows the simulated non-equilibrium measure.
  \textbf{b.} Latent feature representations visualized with UMAP before domain adaptation (left) and after contrastive domain adaptation (right). Cyan markers indicate theoretical XPCS samples, and orange markers indicate experimental XPCS samples.
  \textbf{c.} Prediction results after domain adaptation. Left and middle panel: predicted versus real $D$ and $\lambda_{\mathrm{GB}}$ for simulated test data after adaptation, with a test-set $R^2=0.950$ and $0.890$, respectively. Right panel: inferred experimental XPCS locations on the same theoretical $(D,\lambda_{\mathrm{GB}})$ map, showing their shift into the physically expected crossover region.}
  \label{fig4}
\end{figure}

With both continuum simulation and experimental XPCS data available, our goal is to place each experimental observation onto the theoretical parameter manifold and infer the underlying physical parameters $(D,\Gamma,\lambda_{\mathrm{GB}})$ of each sample at a given temperature.
A natural starting point is a purely supervised regression model trained on simulated data. Specifically, we train a convolutional neural network (CNN) on the 2,000 simulated XPCS spectra to map $g_2(t_1,t_2)$ images to $(D,\Gamma,\lambda_{\mathrm{GB}})$ using a mean-squared-error (MSE) regression loss $\mathcal{L}_y$. We note that in Fig.\,\ref{fig3}c, the crossover regime appears smoother and more well-defined in the $(D, \lambda_\text{GB})$ space compared to other two-parameter projections. Therefore, in this section, we focus on illustrating the $(D, \lambda_{\mathrm{GB}})$ mapping.
As shown in Fig.\,\ref{fig4}a (left and middle panels), this model achieves excellent performance on the simulated data, with a test-set $R^2=0.995$ and $R^2=0.893$ for the predicted parameters $D$ and $\lambda_\mathrm{GB}$ relative to the ground-truth values, indicating that the mapping from XPCS features to physical parameters is well learned within the theoretical domain.
However, this model completely fails when applied directly to experimental data. As is shown in Fig.\,\ref{fig4}a (right panel), when experimental maps are passed through the simulation-trained predictor, the inferred parameters unanimously collapse to outlier regions (e.g., $\lambda_{\mathrm{GB}}\approx0$). Such observation from the vanilla model's output is inconsistent with the smooth crossover structure in $(D,\Gamma)$ identified from theory (Fig.\,\ref{fig3}). This mismatch directly reflects the simulation-to-experiment domain gap. Further tests on data augmentation indicate that the domain gap in this particular scenario is non-trivial, i.e., it can't be fixed by simple operations like smearing or rescaling on the experimental XPCS data (see Supplementary Information).

To address this, we introduce a domain adaptation objective $\mathcal{L}_d$ in addition to the regression loss $\mathcal{L}_y$. The domain loss $\mathcal{L}_d$ imposes two constraints: (i) alignment of feature distributions between simulated and experimental data in the learned representation space, implemented via a CORAL loss \cite{sun2016deep}; and (ii) consistency between predicted parameters and experimentally measured non-equilibrium signatures, enforced during the fine-tuning stage for each sample. More details regarding the formulation of training loss and schedule are described in Methods section.
The effect of the domain adaptation is shown in Fig.\,\ref{fig4}b. We visualize the learned feature space for experimental and simulated data, using the uniform manifold approximation and projection (UMAP) \cite{mcinnes2018umap} for dimension reduction on the feature vectors extracted from CNN. For the vanilla model trained solely on the continuum model's simulation, the simulated and experimental data occupy distinctively separated regions. However, after training with domain loss $\mathcal{L}_d$ on 2,000 simulated and 63 experimental XPCS spectra, the two domains significantly overlap in the two-dimensional UMAP embedding, indicating successful alignment of representations.
Importantly, this alignment does not come at the cost of predictive accuracy on the kinetic parameters in theoretical space. As shown in Fig.\,\ref{fig4}c (left panel), after training the model with domain loss, the predictor retains high test-set $R^2=0.950$ and $R^2=0.890$ on simulated data with well-behaved adaptation, demonstrating that the physical mapping learned from theory is preserved.
With this adapted ML model, we then infer parameters for experimental XPCS maps (which remain unlabeled), thus completing a semi-supervised learning pipeline. The resulting assignments (Fig.\,\ref{fig4}c, right panel) now lie along the physically expected crossover regime in the $(D,\lambda_\text{GB})$ diagram, with trends that are consistent with the continuum model and directly interpretable in terms of competing diffusion and GB dynamics.
In this way, although experimental data are scarce and not directly labeled, they instead act as constraints that select physically meaningful regions of the theoretical manifold, enabling robust and interpretable parameter inference for the non-equilibrium dynamics of GB.

\section*{Discussion}
Our results establish XPCS as a powerful and direct probe of slow, non-equilibrium GB dynamics over long timescales (minutes to hours), in contrast to prior approaches based on discrete real-space snapshots \cite{qiu2024grain}.
By resolving the full two-time correlation function $g_2(t_1,t_2)$, this work moves beyond static or time-averaged descriptions and directly quantifies the breakdown of time-translation invariance through a non-equilibrium metric $\mathcal{S}_\text{noneq}$. 
A key observation is that a minimal continuum model, incorporating the competition between bulk diffusion and curvature-driven GB dynamics, is sufficient to reproduce the essential features of the experimental XPCS maps. Despite its simplicity, this model captures the crossover between equilibrium and non-equilibrium regimes and provides a low-dimensional, physically interpretable parameterization in terms of the bulk diffusivity $D$, the GB stiffness $\Gamma$ and the effective GB concentration $\lambda_{\mathrm{GB}}$. 
Here, $D$ sets the bulk relaxation rate, while $\Gamma$ and $\lambda_\mathrm{GB}$ describe the effective GB relaxation strength and GB-associated dynamic weight, respectively, with $\lambda_\mathrm{GB}$ expected to increase with GB density and decrease with grain size. This indicates that the dominant physics of GB relaxation shown in the experimental data can be effectively recovered without resorting to time-costly atomic-scale simulation.
The primary challenge, however, lies in \textit{quantitatively} connecting the continuum theoretical model to experimental observations. As demonstrated by the failure of the vanilla ML model in Fig.\,\ref{fig4}a, models trained solely on simulated data do not generalize to experiment due to the substantial domain gap arising from noise, instrumental effects, and unmodeled microstructural complexity. 
Our semi-supervised domain adaptation framework addresses this challenge by utilizing both labeled simulated data and unlabeled experimental data, where experimental measurements do not act as supervision in the conventional sense, but instead restrict the model to physically plausible regions of the theoretical parameter manifold. This enables robust and interpretable parameter inference while preserving the predictive accuracy learned from XPCS simulation.

Recent advances in AI for physics and materials engineering have increasingly emphasized the need for a closed loop between AI, computation and experiment \cite{carrasquilla2020machine,chen2021machine,cheng2024machine,han2025ai,fu2025ai,cheng2026artificial,CHENG2026102728}. In this context, the domain gap between theory and experiment has become a central challenge across all subfields.
Several works have attempted to bridge this gap using deep learning. For instance, domain-adversarial learning has been used to extract superconducting phase information from angle-resolved photoemission spectroscopy (ARPES) spectra by aligning simulated and experimental feature representations \cite{chen2025detecting}, while adversarial distribution alignment frameworks train generative models on simulations and then match them to  distributions of experimental observables, such as nuclear magnetic resonance (NMR) spectroscopy, via a min-max objective with Kullback-Leibler (KL) regularization \cite{nelson2026bridging}. 
However, these approaches rely on min-max optimization, which are prone to instability, sensitive to objective design, and challenging to train in regimes with limited experimental data. In this work, we find that a simple CORAL loss based feature alignment, combined with a physically motivated non-equilibrium consistency constraint, provides a stable and effective solution to the XPCS problem. By avoiding adversarial min-max optimization and relying on feature-space convex alignment losses, the framework achieves robust training, even in data-scarce experimental regimes. Our results propose an alternative route to bridge the domain gap between theory and experiment using deep learning.

Despite these advances, several limitations of the present framework should be noted. 
First, the continuum model provides a coarse-grained description of GB and does not explicitly capture atomistic mechanisms such as interactions between multiple grains and dynamic heterogeneity \cite{nye2018direct,zhang2021equation,rohrer2023grain}. Second, the domain adaptation relies on a scalar non-equilibrium metric, which may not fully capture the richness of the underlying dynamics encoded in the two-time correlation function. Third, the approach assumes that experimental observations lie within the span of the simulated parameter space, which may not hold in more complex materials systems or under extreme conditions (see minor outlier examples in Supplementary Information). Addressing these limitations will require both improved physical modeling and more expressive data representations.
For example, on the physics side, incorporating additional degrees of freedom into the continuum model, such as structural disorder, anisotropy, or multi-boundary interactions, could further improve the fidelity of the simulation; on the experimental side, combining XPCS with complementary modalities may help break degeneracies in interpreting non-equilibrium dynamics.

Looking forward, our results demonstrate that combining XPCS, minimal physical modeling, and domain-adaptive ML enables quantitative and interpretable access to non-equilibrium defect dynamics.
More broadly, the integration of experiment, theory, and machine learning suggests a pathway toward a closed-loop framework, in which experimental data iteratively refine theoretical models, and theory guides the interpretation and acquisition of new measurements.
Our approach charts a concrete roadmap to bridging the gap between theory and experiment in complex physical systems by data-driven domain adaptation, even when they are separated by indirect measurement processes and modeling approximations.

\section*{Methods}
\subsection*{Data simulation}
Since we assume that all atoms not affected by the GB are at equilibrium, their dynamics can be described by thermal Browninan motion, i.e. 
\begin{equation}
\left\langle\left[\mathbf{r}_i(0)-\mathbf{r}_i(t)\right]^2\right\rangle_{\mathrm{eq}}=6D t
\end{equation}
which results in the equilibrium contribution in Eq.\,\ref{equilibrium}. 
As for the motion of atoms on GB\cite{chen2018atomistic}, the velocity can be expressed as $\mathbf{v}(\mathbf{r},t)=M\mathbf{F}(\mathbf{r},t)$ in the continuum limit. Furthermore, the force $\mathbf{F}(\mathbf{r},t)$ consists of both a curvature restoring force $\Gamma \sum_{i}h_{ii}$ under the small slope approximation, where $h_{ii}$ is the local $i$-th principal curvature of GB and only $h_{xx}\approx\frac{\partial^2 z}{\partial x^2}$ is nonzero, and thermal noise $\xi(\mathbf{r},t)$ satisfying $\left\langle\xi(\mathbf{r}, t) \xi\left(\mathbf{r}^{\prime}, t^{\prime}\right)\right\rangle=\frac{2 k_B T}{M} \delta\left(\mathbf{r}-\mathbf{r}^{\prime}\right)\delta\left(t-t^{\prime}\right)$. Putting these together we arrive at Eq.\,\ref{EOM}, which can be readily discretized and solved numerically given appropriate initial and boundary conditions. For the numerical simulation, we set
\begin{equation}
z(x=0)=z(x=L)=0; \quad z(t=0)=A\frac{x(L-x)}{L^2}.
\end{equation}
where $L$ is the system size and $A \neq 0$ marks that the GB is initially not under equilibrium shape, i.e. it will start oscillating. With these conditions and EOM, we can simulate the time-dependent evolution of GB, compute the GB contribution to the intermediate scattering function $\langle e^{i\mathbf{q}\cdot (\mathbf{r}_i(t)-\mathbf{r}_j(t'))}\rangle_{\mathrm{GB}}$, and finally produce the simulated XPCS signal in Eq. \ref{g2} combined with Eq. \ref{equilibrium}. For the simulation ensemble used in Fig.\,\ref{fig3} and for supervised model training, we generated 2,000 parameter sets with $\Gamma$ sampled linearly from $1 \times10^{18}~\mathrm{s} \cdot\mathrm{cm}^{-2}$ to $6\times10^{18}~\mathrm{s} \cdot\mathrm{cm}^{-2}$, $D$ sampled logarithmically from $3\times10^{-24}~\mathrm{cm}^{2}\cdot\mathrm{s}^{-1}$ to $3\times10^{-21}~\mathrm{cm}^{2}\cdot\mathrm{s}^{-1}$, $\lambda_{\mathrm{GB}}$ sampled linearly from 0 to 0.5, and $T$ sampled linearly from 300 to 500 K.

\subsection*{Semi-supervised learning with domain adaptation}
We train the parameter inference model with three sub-components: vanilla supervised training on simulations, global CORAL-surrogate adaptation to experimental data, and sample-wise fine-tuning. The network contains two parts, as shown in Fig.\,\ref{fig1}: a convolutional feature extractor $F_\theta$ with parameters $\theta$, mapping the XPCS spectra $g_2(t_1,t_2)$ ($0\le t_1, t_2 \le 2,500$ s) to a feature vector $h\in\mathbb{R}^{128}$; a multilayer-perceptron (MLP) predictor $P_\phi$ with parameters $\phi$, mapping the feature vector to predicted parameters. Given an input two-time correlation map $x$, the model predicts physical parameters
\begin{equation}
\hat{y}=P_\phi(F_\theta(x)),
\quad
y=(D,\Gamma,\lambda_{\mathrm{GB}}).
\end{equation}
The vanilla model is trained only on simulated data, for which the parameter labels are known. Its supervised regression loss is
\begin{equation}
\mathcal{L}_{y}=\frac{1}{N_s}\sum_{i=1}^{N_s}\left|P_\phi(F_\theta(x_i^s))-y_i^s\right|^2,
\end{equation}
where $N_s = 2,000$ is the total number of simulated data, $x_i^s$, $y_i^s$ denote the simulated $g_2$ maps and their corresponding parameters. In practice, all parameters are normalized to $[0,1]$ either linearly (for $\Gamma$ and $\lambda_\mathrm{GB}$) or logarithmically (for $D$).

To adapt the simulation-trained model to the experiment domain, we use the unlabeled experimental maps through feature alignment and non-equilibrium consistency. The total domain loss $\mathcal{L}_d$ is comprised of two parts: the first part, feature alignment loss, is quantified using the CORAL loss \cite{sun2016deep}
\begin{equation}
\mathcal{L}_{\mathrm{CORAL}}=
\left\|\mu_s-\mu_e\right\|^2+
\frac{1}{4d^2}\left\|C_s-C_e\right\|_F^2,
\end{equation}
where $\mu_{s,e} = \mathbb{E}_{i\in\mathcal{D}_\mathrm{batch}}[F_\theta (x_i^{s, e})]$ and $C_{s,e} = \mathbb{E}_{i\in\mathcal{D}_\mathrm{batch}}[F_\theta (x_i^{s, e})F_\theta (x_i^{s, e})^\top]$ are the minibatch feature means and covariance matrices for the simulated and experimental domains, $d=128$ is the feature dimension, and $\left\|\cdot\right\|_F$ is the Frobenius norm. This term aligns the first- and second-order feature statistics directly in the learned representation. The second part, non-equilibrium measure constraint, further uses the measured non-equilibrium measure $\mathcal{S}_{\mathrm{noneq}}$ as a weak experimental constraint. A separate surrogate network $Q_\omega$ is first trained on simulated data to approximate the forward map from normalized parameters to the non-equilibrium measure,
\begin{equation}
Q_\omega(y_i^s)\approx \mathcal{S}_{\mathrm{noneq}}(x_i^s).
\end{equation}
After this pretraining, $Q_\omega$ is frozen. During domain adaptation, the parameters predicted for experimental maps are required to reproduce the measured experimental non-equilibrium measures:
\begin{equation}
\mathcal{L}_{\mathrm{sur}}=
\frac{1}{N_e}\sum_{j=1}^{N_e}
\left|Q_\omega(P_\phi(F_\theta(x_j^e)))-\mathcal{S}_{\mathrm{noneq}}(x_j^e)\right|^2 .
\end{equation}
Overall, the domain loss is given by $\mathcal{L}_d=w_{\mathrm{CORAL}}\mathcal{L}_{\mathrm{CORAL}}+w_{\mathrm{sur}}\mathcal{L}_{\mathrm{sur}}$.
where we set the hyperparameters $w_{\mathrm{CORAL}}=1.0$ and $w_{\mathrm{sur}}=1.2$ in our training. The full objective for semi-supervised learning with domain adaptation is 
\begin{equation}
\mathcal{L}=\mathcal{L}_{y}+\mathcal{L}_d.
\end{equation}
The model is optimized using Adam with learning rate $\eta=3\times10^{-4}$, while the non-equilibrium surrogate is pretrained using Adam with learning rate $10^{-3}$.

Finally, we fine-tune the globally adapted model separately for each material-dose group. For this step, the adapted model is copied, the surrogate $Q_\omega$ is kept frozen, and only the selected trainable part of the predictor is updated using $\mathcal{L}_{\mathrm{sur}}$ on the experimental shots from that group. This material-wise fine-tuning accounts for residual sample-specific offsets while keeping the inferred parameters tied to the simulation-derived non-equilibrium manifold.

\subsection*{XPCS measurements}
X-ray photon correlation spectroscopy (XPCS) measurements were performed at the Coherent Hard X-ray (CHX, 11-ID) beamline at NSLS-II, Brookhaven National Laboratory. A partially coherent beam with energy 9.65~keV ($\lambda = 0.1285$~nm) was selected using a Si(111) double-crystal monochromator. Scattering patterns were collected using a Dectris Eiger X 4M detector positioned 10.08~m downstream. Measurements were carried out in a grazing-incidence reflection geometry with an incidence angle of $0.25^\circ$.

\section*{Acknowledgments}
The authors thank DJ Srolovitz for the helpful discussions. MC acknowledges support from U.S. Department of Energy (DOE), Office of Science (SC), Basic Energy Sciences Award No. DE-SC0020148. BY thanks support from National Science
Foundation (NSF) ITE-2345084. MC, DP, YC, JJT, and ML thank the support from DOE Genesis Mission Multimodal AI for 2D Quantum Magnets (MAIQMag) project. This research used resources at the CHX beamline of the National Synchrotron Light Source II, a U.S. Department of Energy (DOE) Office of Science User Facility operated for the DOE Office of Science by Brookhaven National Laboratory under Contract No. DE-SC0012704. ML acknowledges the support from With support from the Future Energy Systems Center (FUEC) through the MIT Energy Initiative and support from R. Wachnik.

\bibliography{refs.bib} 
\end{document}